\documentclass{webofc}
\usepackage[varg]{txfonts}   % Web of Conferences font
\usepackage{color}
\usepackage[dvipsnames]{xcolor}
\usepackage{caption}
\usepackage{subcaption}
\usepackage{pifont}
\usepackage{hyperref}
\begin{document}
\title{Applications of Lattice Gauge Equivariant Neural Networks}

\author{\firstname{Matteo} \lastname{Favoni}\inst{1,2}\fnsep\thanks{\email{favoni@hep.itp.tuwien.ac.at}} \and \firstname{Andreas} \lastname{Ipp}\inst{1}\fnsep\thanks{\email{ipp@hep.itp.tuwien.ac.at}} \and \firstname{David I.} \lastname{M\"uller}\inst{1}\fnsep\thanks{\email{dmueller@hep.itp.tuwien.ac.at}}}

\institute{Institute for Theoretical Physics, TU Wien, \\
	Wiedner Hauptstr. 8-10, 1040 Vienna, Austria
\and Speaker and corresponding author}

\abstract{%
The introduction of relevant physical information into neural network architectures has become a widely used and successful strategy for improving their performance. In lattice gauge theories, such information can be identified with gauge symmetries, which are incorporated into the network layers of our recently proposed Lattice Gauge Equivariant Convolutional Neural Networks (L-CNNs). L-CNNs can generalize better to differently sized lattices than traditional neural networks and are by construction equivariant under lattice gauge transformations.
In these proceedings, we present our progress on possible applications of L-CNNs to Wilson flow or continuous normalizing flow. Our methods are based on neural ordinary differential equations which allow us to modify link configurations in a gauge equivariant manner. For simplicity, we focus on simple toy models to test these ideas in practice.
}
\maketitle
\section{Introduction}

In the past decade, neural networks (NNs) have been established as an essential tool with numerous applications in e.g.~computer science and the natural sciences. The crucial role played by symmetries in a large amount of scientific problems has attracted the idea that the inclusion of such symmetries in the NN architecture could be beneficial in enhancing their performance. For example,  convolutional neural networks (CNNs) are based on the inclusion of translational symmetry as an inherent property of their architecture. In computer vision problems such as image classification, this proved to be a powerful idea, due to the fact that the position of a particular feature that has to be detected is irrelevant and it is just the presence of the feature that matters. This approach has been generalized to include other symmetries in group equivariant convolutional neural networks (G-CNNs)~\cite{Cohen:2016}, which take into account not just translational, but e.g.~also rotational and reflection symmetry. Recently, this idea has been further extended to local symmetries~\cite{Cohen:2019}. The more general framework dealing with symmetries in neural networks is called geometric deep learning \cite{Gerken:2021sla}.

In theoretical physics, and more specifically in lattice field theories with global symmetries, CNNs or G-CNNs have been successfully applied to solving regression problems and detecting phase transitions \cite{Zhou:2019, Boyda:2020nfh, Blucher:2020mjt, Bachtis:2020ajb, Bulusu:2021rqz, Bachtis:2021xoh, Bachtis:2021eww}. In the context of Abelian and non-Abelian gauge theories, which exhibit local symmetries, there has been progress in the direction of incorporating gauge symmetry in the network architecture \cite{Kanwar:2020, Boyda:2020, Tomiya:2021ywc, Luo:2020stn, Abbott:2022zhs}. For example, gauge equivariant normalizing flows~\cite{Kanwar:2020,Boyda:2020, Albergo:2021vyo,Abbott:2022zhs} can be used in place of Monte Carlo simulations to sample uncorrelated gauge configurations while retaining gauge symmetry. Similarly, a lattice gauge equivariant convolutional neural network (L-CNN) was proposed in our paper~\cite{Favoni:2020}, in which the elementary layers of the architecture individually preserve gauge symmetry. L-CNNs have been used successfully for regression tasks and in principle can also be employed for the generation of gauge configurations. The continuous flow approach proposed in~\cite{deHaan:2021erb,Gerdes:2022eve} provides a continuous generalization of normalizing flows applied to lattice field theory. In contrast to normalizing flows, this continuous formulation allows for a straightforward inclusion of exact global symmetries. At its core, continuous flows are an application of neural ordinary differential equations (NODEs) \cite{Chen:2018_69386f6b}, which are ordinary differential equations (ODEs) parametrized by NNs.

In these proceedings, we first review the basics of lattice gauge theory and the L-CNN architecture, and we show how NODEs can be modified to study Wilson flow \cite{Luscher:2010iy} and exemplify it with an SU(2) toy model.

\section{Lattice gauge equivariant neural networks}

\begin{figure}[t]
    \centering
    \includegraphics[width=\textwidth]{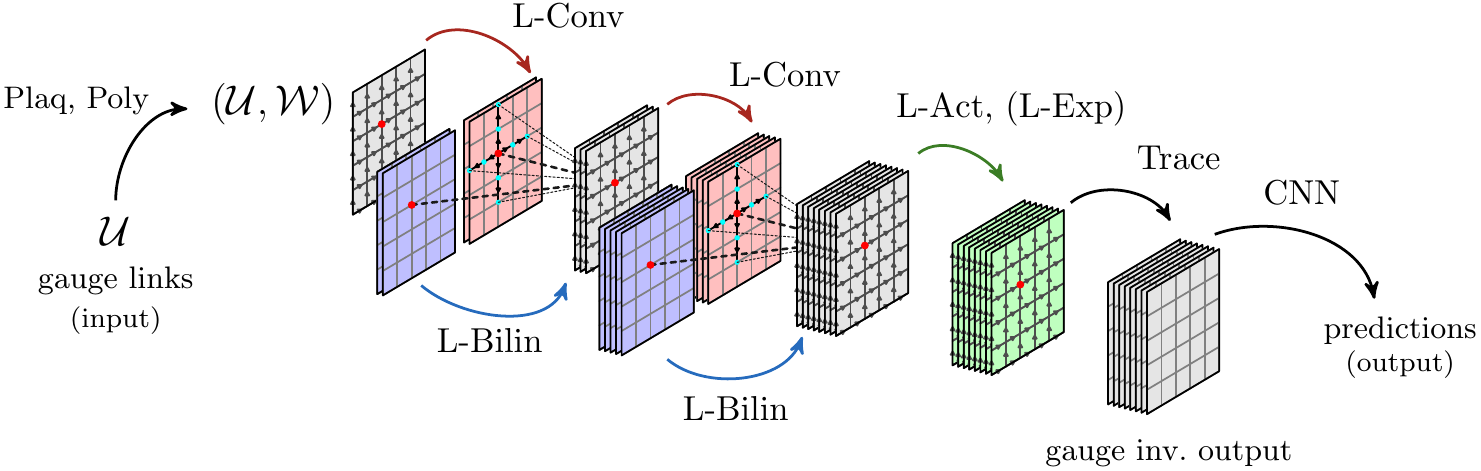}
    \caption{A possible L-CNN architecture. Figure from~\cite{Favoni:2020}}
    \label{fig:lcnn}
\end{figure}

Lattice gauge theory is a discretized version of $\text{SU}(N_c)$ Yang-Mills theory, in which spacetime is approximated by a periodic hypercubic lattice in $D+1$ dimensions with imaginary time and with lattice spacing $a$. In the lattice discretizations, the continuous gauge fields $A_\mu$ are replaced by the link variables $U_{x,\mu}$ via the following definition:
\begin{equation}
    U_{x,\mu}=\mathcal{P}\exp\left(ig\int_x^{x+\hat\mu}\mathrm{d}x'^\nu A_\nu(x')\right),
\end{equation}
where $\mathcal{P}$ denotes path ordering, $g$ is the coupling constant and the integral is performed over the straight line connecting the site $x$ to the site $x+\hat{\mu}$.
The gauge fields are elements of the $\mathfrak{su}(N_c)$ algebra, while the links can be interpreted as the parallel transporters along the lattice edges and live in the $\text{SU}(N_c)$ group. In practice, we employ the fundamental representation of $U_{x,\mu}$, where links are represented as complex $N_c \times N_c$ matrices.

It is possible to multiply adjacent links, the repetition of which leads to arbitrary Wilson lines. If the start and end point of a Wilson line coincide, closed loops are formed and are called Wilson loops. The simplest Wilson loop on a hypercubic lattice is the plaquette given by
\begin{equation}
    U_{x,\mu\nu}=U_{x,\mu}U_{x+\hat{\mu},\nu}U_{x+\hat{\mu}+\hat{\nu},\nu}^\dagger U_{x+\hat{\nu},\mu}^\dagger.
\end{equation}

The Wilson action~\cite{Wilson:1974sk}, formulated in terms of plaquettes,
\begin{equation}
    S_W[U] = \frac{2}{g^2} \sum_{x \in \Lambda} \sum_{\mu < \nu} \mathrm{Re} \, \mathrm{Tr} \left[ \mathbf{1} - U_{x,\mu\nu} \right],
\end{equation}
is equivalent to the Yang-Mills action in the continuum limit $a\rightarrow0$. A general lattice gauge transformation applied to links
\begin{equation}
    U_{x,\mu}\rightarrow\Omega_x U_{x,\mu}\Omega_{x+\hat{\mu}}^\dagger
\end{equation}
induces a local transformation of the plaquettes
\begin{equation}
    U_{x,\mu\nu}\rightarrow\Omega_x U_{x,\mu\nu}\Omega_x^\dagger.
\end{equation}
These transformations leave the Wilson action unchanged, meaning the theory is invariant under $\text{SU}(N_c)$ lattice gauge transformations.

In order to build up an L-CNN~\cite{Favoni:2020}, we can consider its individual layers, each of which is designed to respect gauge equivariance. First, the input consists of the set of gauge links $\mathcal{U}$ of a particular lattice configuration and locally transforming objects $\mathcal{W}$, which in practice we choose to be the plaquettes, but can also be the Polyakov loops (closed Wilson lines wrapping around the periodic boundary of the lattice). The L-Conv layer is a gauge equivariant convolution, which acts as a parallel transporter of locally transforming objects $\mathcal{W}$, while L-Bilin performs a multiplication of such objects (more specifically, it is a bilinear layer). We proved that the repeated application of these two operations can grow arbitrarily sized Wilson (or Polyakov) loops. Moreover, it is possible to introduce non-linearity via L-Act layers, which behave like activation functions in traditional CNNs. The Trace layer yields a gauge invariant output that can be passed to a traditional CNN. A possible realization of such a network is depicted in Fig.~\ref{fig:lcnn}. By virtue of the ability of generating any loop and the non-linearity, L-CNNs can be seen as universal approximators of gauge-equivariant functions on the lattice.

Among the relevant results found in~\cite{Favoni:2020}, it is worth mentioning that the L-CNNs performed very well on the regression of Wilson loops up to a size of $4\times4$ and simple observables such as the topological charge density, beating traditional CNNs on the same task.

\section{Adaptation of NODEs to lattice gauge theory}

NODEs are ODEs parametrized by neural networks~\cite{Chen:2018_69386f6b}. As in the original paper, we will focus on first order ODEs:
\begin{equation}
\label{node}
    \frac{\mathrm{d}\mathbf{z}}{\mathrm{d}t}=\mathbf{f}(\mathbf{z}(t),\theta,t).
\end{equation}
The unknown function $\mathbf{z}(t)$ is a time-dependent $D$-dimensional vector and $\mathbf{f}(\mathbf{z}(t),\theta,t)$ is a $D$-dimensional function  parametrized by a priori unknown weights $\theta$. In particular one can choose $\mathbf{f}(\mathbf{z}(t),\theta,t)$ to be represented by a NN. NODEs can be understood as generalizations of residual networks~\cite{He:2015} with continuous depth, where the time coordinate $t$ is used in place of the discrete depth of the network. Starting with an input state  $\mathbf{z}_0=\mathbf{z}(t_0)$ at $t=t_0$, the NODE can be formally solved by
\begin{align}
    \mathbf{z}(t_1)=\mathbf{z}_0+\int_{t_0}^{t_1}\mathrm{d}t'\,\mathbf{f}(\mathbf{z}(t'),\theta,t'),
\end{align}
which provides predicted states $\mathbf{z}(t_1)$ at some final time $t=t_1$. In this manner, NODEs map arbitrary input states to output states similar to generic NNs. The mapping depends on the NN architecture and the weights $\theta$. NODEs can thus be used to solve regression problems: given a dataset characterized by the initial conditions $\mathbf{z}_0^i=\mathbf{z}^i(t_0)$ (where $i\in\{1,\dots,N_{\text{samples}}\}$), which are used as input, and the desired output vectors $\mathbf{\tilde{z}}_1^i$, which represent the labels, the weights $\theta$ can be optimized such that the final states approximate the labels as accurately as possible. In practice, this is done with the aid of an ODE integrator, such as Euler or Runge-Kutta. We can require that the discrepancy between the labels and the predicted final states is minimized by introducing a loss function such as the mean squared error (MSE), $\mathcal{L}(\theta)=\sum_i(\mathbf{\tilde{z}}_1^i-\mathbf{z}^i(t_1))^2/N_{\text{samples}}$, and run the training procedure in order to optimize the weights $\theta$.

While the approach described above only uses the final state labels in the optimization problem, it is also possible to include the discrepancies of the whole state evolution (i.e.~more points $t_j$ along the trajectory $\mathbf z(t_j)$) in the loss function for successful training. If we only use the final states at $t_1$, then it is crucial that the dataset provides sufficient information to reconstruct the underlying dynamics.

We can adapt the previous scheme to study continuous flow applied to lattice gauge configurations
\begin{equation}
\label{eq_flow}
    \frac{\mathrm{d}U_{x,\mu}(\tau)}{\mathrm{d}\tau}=iH_\mu[U_{x,\mu}(\tau),\theta,\tau]\,U_{x,\mu}(\tau),
\end{equation}
where $U_{x,\mu}\in\text{SU}(N_c)$ are gauge links, $\tau$ is flow time and
$H_\mu[U(\tau),\theta,\tau]$ is a NN parametrized by the weights $\theta$ with a traceless and Hermitian output. This last requirement guarantees that the gauge links do not leave the group during the evolution. In order to retain gauge equivariance, $H_\mu$ can be modeled with an L-CNN.  

Our dataset consists of the initial conditions $U_{x,\mu,0}^i=U_{x,\mu}^i(\tau_0)$ and the desired output configurations $\tilde{U}_{x,\mu,1}^i$, which define input and labels respectively. A standard ODE integrator would in general break the group structure, so we make use of the iterative application of the exponential map
\begin{align}
   U_{x,\mu}^i(\tau_{j+1})=\exp\left(iH_\mu[U^i(\tau_j), \theta, \tau_j]\Delta \tau\right)U_{x,\mu}^i(\tau_j)
\end{align}
for time evolution. Since $H_\mu$ is traceless and Hermitian, i.e.~it can be understood as a $\mathfrak{su}(N_c)$ algebra element, the links remain in the group manifold. The final configuration is then used in a loss function, such as 
\begin{align}
    \mathcal{L}(\theta)=\frac{1}{N_{\text{samples}}} \sum_{x,\mu}\sum_i\|\tilde{U}_{x,\mu,1}-U_{x,\mu}^i(\tau_1)\|^2,
\end{align}
where $\| \dots \|$ denotes the Frobenius norm.

\section{SU(2) Wilson flow toy model}

\begin{figure}[t]
    \centering
    \includegraphics[width=\textwidth]{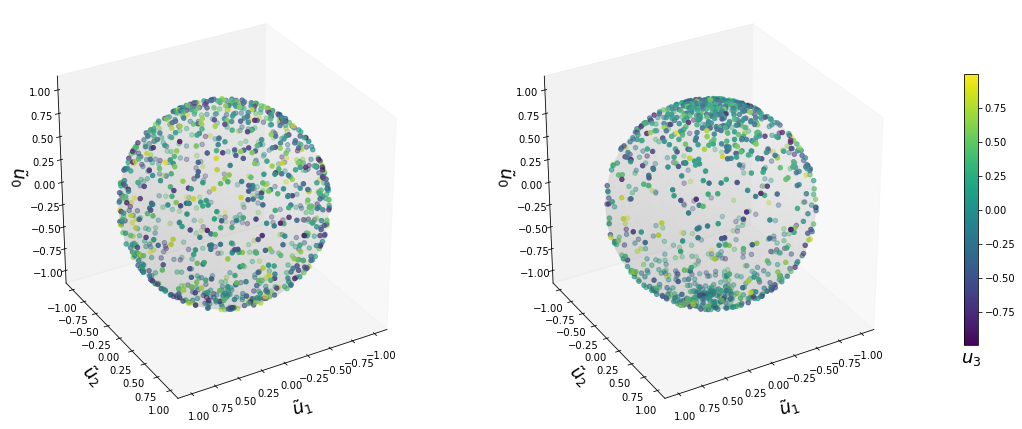}
    \caption{Visualization of 1000 samples from the dataset. Left: distribution of the initial conditions $U_0^i$. Right: distribution of the labels $\tilde{U}_1^i$, found by evolving the initial conditions according to the action $S[U]=\mathrm{Re}\,\mathrm{Tr}\, (U^2)$ up to $\tau_1=1$.}
    \label{fig:dataset}
\end{figure}

We test the adaptation of the NODE approach to Wilson flow  \cite{Luscher:2010iy} using a toy model consisting of a single $\text{SU}(2)$ link characterized by the action $S[U]=\mathrm{Re}\,\mathrm{Tr}\, (U^2)$. Starting with randomly distributed initial matrices $U^i_0 \in \text{SU}(2)$, we generate a dataset of flowed matrices $\tilde U^i_1$ by applying gradient descent on $S[U]$ using group derivatives akin to Wilson flow in lattice gauge theory. The action exhibits two minima, $\pm \mathbf{1}$, toward which Wilson flow lets the links evolve depending on the initial conditions. If $\mathrm{Tr}\,U>0$, the link is flowed toward the north pole ($+\mathbf{1}$), otherwise the dynamics is directed toward the south pole ($-\mathbf{1}$). For links with $\mathrm{Tr} U = 0$, the dynamics are stuck. Our goal is to use NODEs to reconstruct these dynamics via the flow Eq.~\eqref{eq_flow}. 

In order to visualize the dataset, we use the following parametrization of $\text{SU}(2)$
\begin{equation}
    U=u_0\mathbf{1}+i\sigma_au_a, \qquad u_0=\frac{1}{2}\mathrm{Tr}\,(U), \qquad u_i=\frac{1}{2i}\mathrm{Tr}\,(U\sigma_i),
\end{equation}
where $\sigma_i$ are the Pauli matrices. We then normalize $u_0$, $u_1$ and $u_2$ by introducing ${\Tilde{u}_j=u_j/\sqrt{u_0^2+u_1^2+u_2^2}}$ for $j\in\{0,1,2\}$, so that each $\tilde{u}_j$ lies on a three-dimensional sphere, while the remaining parameter $u_3$ determines the color of a point, as shown in Fig.~\ref{fig:dataset}. As anticipated, the link variables, which are initially homogeneously distributed on the sphere, flow toward one of the two minima, which in Fig.~\ref{fig:dataset} correspond to the north and south pole of the sphere.

Training NODEs in principle needs backpropagation through the ODE solver. The problem with standard backpropagation is that it requires to keep the whole evolution of the system, which leads to increased memory consumption. For more complicated systems, this can easily saturate memory. A solution to this problem lies in the adjoint sensitivity method \cite{Chen:2018_69386f6b}, which avoids having to store the entire history of the Wilson flow by solving the evolution backwards. Our implementation of this method is still a work in progress, so for this simple system we rely on standard backpropagation.

For our model, the matrix $H$ in Eq.~\eqref{eq_flow} is constructed with the following steps: the complex entries of $U$ are split into real and imaginary parts. They are then fed into a multi-layer perceptron with real weights. The output is recombined into a complex matrix, which is generally neither Hermitian nor traceless. Therefore, we take its anti-hermitian traceless part, $\left[ C \right]_\mathrm{ah} = \left(C \! - \! C^\dagger \right) / (2 i) - \mathbf{1} \, \mathrm{Tr} \left( C \! - \! C^\dagger\right) / (2 i N_c)$, which projects the output onto the $\mathfrak{su}(2)$ algebra. The application of the exponential map yields a matrix in SU(2).  This guarantees that the evolution of $U$ takes place without leaving the group.

We choose the Frobenius norm averaged over the lattice as our loss function and train on a dataset of 50000 samples using a batch size of 100 and a learning rate of $10^{-3}$ for 100 epochs. The multi-layer perceptron we employed has four hidden layers with 16, 64, 32 and 16 nodes respectively. We use $\tanh(x)$ as an activation function after every layer except the last.

\begin{figure}[t]
    \minipage{0.5\textwidth}
    \centering
    \begin{subfigure}{\textwidth}
    \includegraphics[width=\textwidth]{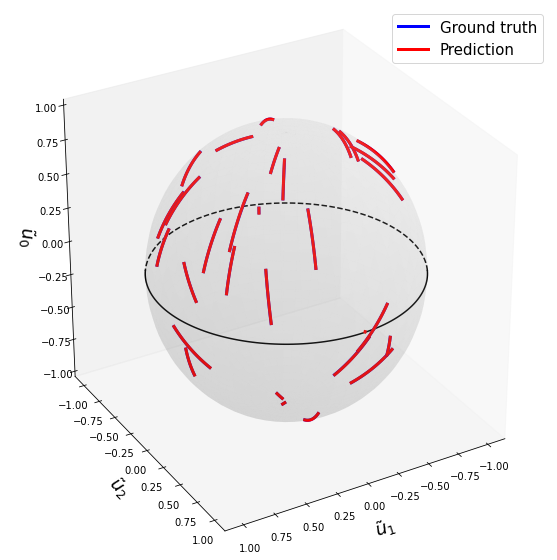}
    \caption{Ground truth and predicted trajectories}
    \end{subfigure}
    \endminipage
    \hfill
    \minipage{0.5\textwidth}
    \centering
    \begin{subfigure}{\textwidth}
    \includegraphics[width=.9\textwidth]{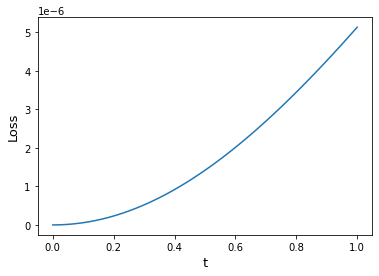}
    \caption{Loss as a function of flow time}
    \end{subfigure}
    \endminipage
    \caption{Test results. (a) Evolution of 30 samples projected on the three-dimensional sphere. The ground truth and the prediction lie on top of each other. (b) The corresponding Frobenius loss as a function of flow time.}
    \label{fig:test_res}
\end{figure}

After training, we test on 4000 samples with the same final Wilson flow time $\tau=1$. The results are shown in Fig.~\ref{fig:test_res}. The left panel shows the trajectories of the ground truth (blue) and the predicted trajectories (red) during the NODE flow. The right panel shows the MSE as a function of flow time. Since the loss of $5\cdot10^{-6}$ is very small, the  two evolutions are visually indistinguishable.

\begin{figure}[t]
    \minipage{0.5\textwidth}
    \centering
    \begin{subfigure}{\textwidth}
    \includegraphics[width=\textwidth]{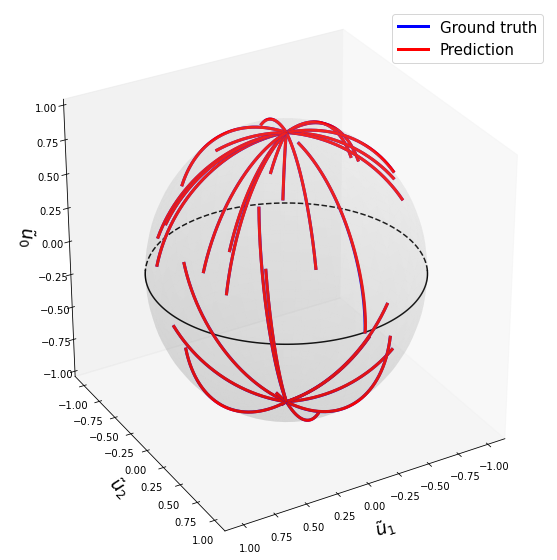}
    \caption{Ground truth and predicted trajectories}
    \end{subfigure}
    \endminipage
    \hfill
    \minipage{0.5\textwidth}
    \centering
    \begin{subfigure}{\textwidth}
    \includegraphics[width=.9\textwidth]{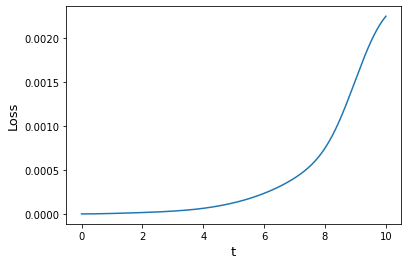}
    \caption{Loss as a function of flow time}
    \label{fig:extrap_res_b}
    \end{subfigure}
    \endminipage
    \caption{Results of our extrapolation up to $\tau=10$ based on training up to $\tau=1$. (a) Extrapolated evolution of 30 samples. The ground truth and the prediction are very close. (b) Loss function as a function of flow time. At large times the loss increases by up to three orders of magnitude compared to the original interval $\tau \in[0, 1]$.}
    \label{fig:extrap_res}
\end{figure}

Since the loss function in Fig.~\ref{fig:test_res} (b) seems to increase quadratically as a function of flow time, we investigate the loss at larger times $\tau > 1$ outside the training interval. In Fig.~\ref{fig:extrap_res} we test 4000 samples and extrapolate to flow times up to $\tau=10$. The deterioration of the performance is clear, since the loss jumps up to values that are three orders of magnitude larger compared to the highest loss found during testing in the interval $\tau \in [0, 1]$. 
Investigating our data more closely, we found two types of mispredictions which can contribute to a large loss. For one specific sample in our test dataset, the predicted trajectory moves in the opposite direction of the actual one. There are also some trajectories at large times $\tau$ that tend to overshoot the ground truth values. In both cases, we found that these trajectories originate from points that lie within a thin neighborhood of the equator ($\mathrm{Tr} U \approx 0$) and are in general very difficult for the network to evolve correctly. Despite these flaws, the results are encouraging, considering that we employed a simple multi-layer perceptron which is not adapted to the symmetries of the problem. Therefore, a network structure incorporating additional symmetries ($U \rightarrow \Omega U \Omega^\dagger$ with $\Omega\in \text{SU}(2)$), could further improve the performance in this toy model.

\section{Conclusions and outlook}

In these proceedings, we reviewed the structure of L-CNNs and their successful application in regression tasks. These architectures are very flexible and their layers can be composed to modify gauge link configurations. We discussed how this can be achieved in the context of NODEs and tested it on an $\text{SU}(2)$ Wilson flow toy model with a single link. Based on these experiments, we intend extend the toy model to actual lattice configurations and apply L-CNNs to Wilson flow.

\vspace{1em}

\begin{acknowledgement}
This work has been supported by the Austrian Science Fund FWF No.~P32446, No.~34764 and Doctoral program No.~W1252-N27. The Titan\,V GPU used for this research was donated by the NVIDIA Corporation.
\end{acknowledgement}

\bibliography{references.bib}

\begin{thebibliography}{23}

\bibitem{Cohen:2016}
T.S. Cohen, M.~Welling, \emph{Group Equivariant Convolutional Networks}, in
  \emph{Proceedings of The 33rd International Conference on Machine Learning}
  (JMLR, 2016), Vol.~48, pp. 2990--2999, \texttt{1602.07576}

\bibitem{Cohen:2019}
T.S. Cohen, M.~Weiler, B.~Kicanaoglu, M.~Welling, \emph{Gauge Equivariant
  Convolutional Networks and the Icosahedral {CNN}}, in \emph{Proceedings of
  the 36th International Conference on Machine Learning} (JMLR, 2019), Vol.~97,
  pp. 1321--1330, \texttt{1902.04615}

\bibitem{Gerken:2021sla}
J.E. Gerken, J.~Aronsson, O.~Carlsson, H.~Linander, F.~Ohlsson, C.~Petersson,
  D.~Persson (2021), \texttt{2105.13926}

\bibitem{Zhou:2019}
K.~Zhou, G.~Endr\H{o}di, L.G. Pang, H.~St\"ocker, Phys. Rev. D \textbf{100},
  011501 (2019), \texttt{1810.12879}

\bibitem{Boyda:2020nfh}
D.L. Boyda, M.N. Chernodub, N.V. Gerasimeniuk, V.A. Goy, S.D. Liubimov, A.V.
  Molochkov, Phys. Rev. D \textbf{103}, 014509 (2021), \texttt{2009.10971}

\bibitem{Blucher:2020mjt}
S.~Bl\"ucher, L.~Kades, J.M. Pawlowski, N.~Strodthoff, J.M. Urban, Phys. Rev. D
  \textbf{101}, 094507 (2020), \texttt{2003.01504}

\bibitem{Bachtis:2020ajb}
D.~Bachtis, G.~Aarts, B.~Lucini, Phys. Rev. E \textbf{102}, 053306 (2020),
  \texttt{2007.00355}

\bibitem{Bulusu:2021rqz}
S.~Bulusu, M.~Favoni, A.~Ipp, D.I. M\"uller, D.~Schuh, Phys. Rev. D
  \textbf{104}, 074504 (2021), \texttt{2103.14686}

\bibitem{Bachtis:2021xoh}
D.~Bachtis, G.~Aarts, B.~Lucini, Phys. Rev. D \textbf{103}, 074510 (2021),
  \texttt{2102.09449}

\bibitem{Bachtis:2021eww}
D.~Bachtis, G.~Aarts, F.~Di~Renzo, B.~Lucini, Phys. Rev. Lett. \textbf{128},
  081603 (2022), \texttt{2107.00466}

\bibitem{Kanwar:2020}
G.~Kanwar, M.S. Albergo, D.~Boyda, K.~Cranmer, D.C. Hackett, S.~Racani\`ere,
  D.J. Rezende, P.E. Shanahan, Phys. Rev. Lett. \textbf{125}, 121601 (2020),
  \texttt{2003.06413}

\bibitem{Boyda:2020}
D.~Boyda, G.~Kanwar, S.~Racani\`ere, D.J. Rezende, M.S. Albergo, K.~Cranmer,
  D.C. Hackett, P.E. Shanahan, Phys. Rev. D \textbf{103}, 074504 (2021),
  \texttt{2008.05456}

\bibitem{Tomiya:2021ywc}
A.~Tomiya, Y.~Nagai (2021), \texttt{2103.11965}

\bibitem{Luo:2020stn}
D.~Luo, G.~Carleo, B.K. Clark, J.~Stokes, Phys. Rev. Lett. \textbf{127}, 276402
  (2021), \texttt{2012.05232}

\bibitem{Abbott:2022zhs}
R.~Abbott et~al., Phys. Rev. D \textbf{106}, 074506 (2022), \texttt{2207.08945}

\bibitem{Albergo:2021vyo}
M.S. Albergo, D.~Boyda, D.C. Hackett, G.~Kanwar, K.~Cranmer, S.~Racani\`ere,
  D.J. Rezende, P.E. Shanahan (2021), \texttt{2101.08176}

\bibitem{Favoni:2020}
M.~Favoni, A.~Ipp, D.I. M\"uller, D.~Schuh, Phys. Rev. Lett. \textbf{128},
  032003 (2022), \texttt{2012.12901}

\bibitem{deHaan:2021erb}
P.~de~Haan, C.~Rainone, M.C.N. Cheng, R.~Bondesan (2021), \texttt{2110.02673}

\bibitem{Gerdes:2022eve}
M.~Gerdes, P.~de~Haan, C.~Rainone, R.~Bondesan, M.C.N. Cheng (2022),
  \texttt{2207.00283}

\bibitem{Chen:2018_69386f6b}
R.T.Q. Chen, Y.~Rubanova, J.~Bettencourt, D.K. Duvenaud, \emph{Neural Ordinary
  Differential Equations}, in \emph{Advances in Neural Information Processing
  Systems}, edited by S.~Bengio, H.~Wallach, H.~Larochelle, K.~Grauman,
  N.~Cesa-Bianchi, R.~Garnett (Curran Associates, Inc., 2018), Vol.~31,
  \texttt{1806.07366}

\bibitem{Luscher:2010iy}
M.~L\"uscher, JHEP \textbf{08}, 071 (2010), [Erratum: JHEP 03, 092 (2014)],
  \texttt{1006.4518}

\bibitem{Wilson:1974sk}
K.G. Wilson, Phys. Rev. D \textbf{10}, 2445 (1974)

\bibitem{He:2015}
K.~He, X.~Zhang, S.~Ren, J.~Sun, \emph{Deep Residual Learning for Image
  Recognition}, in \emph{2016 IEEE Conference on Computer Vision and Pattern
  Recognition (CVPR)} (2016), pp. 770--778

\end{thebibliography}

\end{document}